\documentclass[10pt,prl,showpacs,twocolumn,unsortedaddress,nofootinbib,floatfix,showkeys]{revtex4-1}
\usepackage{bm}
\usepackage{physics}
\usepackage[version=4]{mhchem}
\usepackage{mathtools,mathrsfs,amsfonts,dsfont}
\usepackage{relsize}
\usepackage{scalerel}
\usepackage{todonotes}
\usepackage{ulem}
\usepackage{xcolor}
\usepackage[unicode=true, pdfusetitle, bookmarks=true, bookmarksnumbered=false, bookmarksopen=false, breaklinks=true, pdfborder={0 0 0}, backref=false, colorlinks=true, linkcolor=blue, citecolor=blue, urlcolor=blue]{hyperref}
\usepackage{enumitem}
\usepackage{cancel}
\allowdisplaybreaks

\newcommand{\sm}{\kern0.1em}

\newcommand{\Ep}{\sm\skew{3}{\vu}{\bm{\theta}}}

\DeclareMathAlphabet{\mathbbold}{U}{bbold}{m}{n}

\begin{document}

\title{Comment on the ``Significance of Electromagnetic Potentials in the Quantum Theory''}

\author{Siddhant Das}
\email{Siddhant.Das@physik.uni-muenchen.de}
\affiliation{Mathematisches Institut, Ludwig-Maximilians-Universit\"at M\"unchen, Theresienstr.\ 39, D-80333 M\"unchen, Germany}
%\date{May 10, 2021}
\date{April.\ 19, 2022}

\begin{abstract}
A missing angle-dependent prefactor in the wave functions for half-integer magnetic flux obtained in [Y.\ Aharonov and D.\ Bohm, \href{https://doi.org/10.1103/PhysRev.115.485}{Phys.\ Rev.\ \textbf{115}, 485 (1959)}] is supplied, without which the result is not single-valued. No physical conclusions of the paper are affected.
\end{abstract}

\maketitle
\normalem

In their celebrated 1959 paper \cite{AB59}, Aharonov and Bohm (hereafter AB) considered the motion of an electron of mass \(m\) and charge \(-\sm e\) in the field of an infinitely long solenoid of vanishing radius, and governed by the (Coulomb gauge) Hamiltonian
\begin{equation}
    H=\frac{1}{2m}\!\left(-\sm i\hbar\sm\pmb{\grad}+\frac{e\Phi}{2\sm\pi r}\Ep\right)^{\!2}\!,
\end{equation}
where \(\smash{(r,\theta)}\) denote plane-polar coordinates, and \(\Phi\) is the magnetic flux linked with the solenoid. The axis of the solenoid defines the \(z\)-axis. Exploiting the translational symmetry of the problem, AB factor out the motion along the \(z\) direction, obtaining a complete set of eigenfunctions \(F_\alpha(kr,\theta)\) of \(H\)  with eigenvalue \(\smash{E=\hbar^2k^2/(2m)}\). Here,
\begin{equation}\label{psialpha}
    F_\alpha(r,\theta)=\sum_{n=-\infty}^{\infty}(-\sm i)^{|n+\alpha|}J_{|n+\alpha|}(r)\sm e^{in\theta},
\end{equation}
where \(J_\nu(\cdot)\) is Bessel's function of the first kind and \[\alpha=-\sm e\Phi/(2\sm\pi\hbar)\] is the nondimensionalized magnetic flux. The propagator for this problem, i.e., the matrix element of the time-evolution operator\(\sm\matrixel{\vb{r}^\prime}{\exp(-\sm itH/\hbar)}{\vb{r}}\), is directly expressible in terms of \(F_\alpha\):
\begin{equation}
    K\big(\vb{r},\vb{r}^\prime;t\big)=\frac{1}{2\sm\pi i\kern-0.1em \tau}\exp[\frac{i}{2\tau}\big(r^2+r^{\prime\sm2}\big)]\sm F_\alpha\!\left(\frac{rr^\prime}{\tau},\theta-\theta^\prime\right)\!,
\end{equation}
where \(\smash{\tau=\hbar\sm t/ m}\) \cite[Eq.\ (7)]{propagator}. Therefore, it is useful to evaluate \eqref{psialpha} in closed form, say, for special values of \(\alpha\). 

In \cite[p.\ 490]{AB59} AB noticed that \eqref{psialpha} can be summed explicitly for integer and half-integer \(\alpha\)s. In the latter case they present without derivation the result \cite[Eq.\ (23)]{AB59}
\begin{equation}\label{AB23}
    F_{n+1/2}(r,\theta)=\sqrt{\frac{i}{2}}\,\,e^{-\sm i\theta/2-\sm ir\cos\theta}\!\int_0^{\sqrt{r\sm(1+\cos\theta)}}\!\!\!\!dz~e^{iz^2},
\end{equation}
which may be related to the error function \(\text{erf}(\cdot)\). However, this expression is \emph{not} single-valued, as
\[F_{n+1/2}(r,\theta+2\sm\pi)\ne F_{n+1/2}(r,\theta).\]
Thus \eqref{AB23} cannot be a special case of \eqref{psialpha} (which is single-valued regardless of \(\alpha\)). Furthermore, the right-hand side of \eqref{AB23} does not depend on \(n\), the integer-part of \(\alpha\). This seems suspicious since \(H\) depends on \(\Phi\) (which is \(n\)-dependent via the definition of \(\alpha\)), while its eigenfunction \eqref{AB23} appears to be independent of \(n\).

In what follows, we evaluate the infinite series \eqref{psialpha} for half-integer \(\alpha\)s via Laplace transforms, obtaining a result that mainly differs from \eqref{AB23} by a \(\theta\)-dependent prefactor, \emph{viz.},
\(\smash{\text{sgn}\big(\kern-0.1em\cos(\theta/2)\big)\sm e^{-\sm i n\theta}}\). This missing factor correctly restores single-valuedness of AB's result as well as its anticipated \(n\)-dependence.
 
First, let \(n\mapsto n-\lfloor\alpha\rfloor\) in \eqref{psialpha}, where \(\lfloor\alpha\rfloor\) denotes the greatest integer \(\le\alpha\), to obtain
\begin{equation}\label{template1}
    F_\alpha(r,\theta)=e^{-\sm i\lfloor\alpha\rfloor\theta}\!\!\sum_{n=-\infty}^{\infty}(-\sm i)^{|n+\{\alpha\}|}J_{|n+\{\alpha\}|}(r)\sm e^{in\theta}.
\end{equation}
Here, \(\smash{\{\alpha\}=\alpha-\lfloor\alpha\rfloor}\) is the fractional part of \(\alpha\) (\(\smash{0\le\{\alpha\}<1}\) for any \(\alpha\)).

For \(\smash{\{\alpha\}=0}\), i.e., integer \(\alpha\), \eqref{template1} can be evaluated in closed form. For this, note that \(\smash{J_{-n}(r)=(-1)^n\sm J_n(r)}\), and
\begin{equation*}
    (-\sm i)^{|n|}=(-\sm i)^n\begin{cases}\,1, &n\ge0\\ \,(-1)^n, &n<0\end{cases},
\end{equation*}
for any integer \(n\). Thus, invoking the Jacobi-Anger expansion \cite[Sec.\ 8.511.4.]{GH}, we obtain
\begin{align}
   F_\alpha(r,\theta)=e^{-\sm i\lfloor\alpha\rfloor\theta}\!\!\sum_{n=-\infty}^{\infty}\!i^n J_n(r)\sm e^{in(\theta-\pi)}=e^{-\sm i(\lfloor\alpha\rfloor\theta+r\cos\theta)}\!,
\end{align}
in agreement with AB after identifying \(r\cos\theta=x\), see \cite[p.\ 490]{AB59}.

It remains to consider \(\smash{\{\alpha\}>0}\) only, for which
\begin{align}\label{template2}
    F_\alpha(r,\theta)&\overset{\eqref{template1}}{=}e^{-\sm i\lfloor\alpha\rfloor\theta}\!\left[\,\sum_{n=0}^{\infty}(-\sm i)^{n+\{\alpha\}}J_{n+\{\alpha\}}(r)\sm e^{in\theta}\right.\nonumber\\
    &\kern2cm\left.+\sum_{n=-\infty}^{-1}(-\sm i)^{-n-\{\alpha\}}J_{-n-\{\alpha\}}(r)\sm e^{in\theta}\right]\nonumber\\
    &\kern-1.2cm\text{(letting \(n\mapsto-1-n\) in the second sum)}\nonumber\\
    &=e^{-\sm i\lfloor\alpha\rfloor\theta}\Big[(-\sm i)^{\{\alpha\}}f_{\{\alpha\}}\big(r,\theta\big)\nonumber\\
    &\kern1.7cm +e^{-\sm i\theta}(-\sm i)^{1-\{\alpha\}}f_{1-\{\alpha\}}\big(r,-\sm\theta\big)\Big],
\end{align}
where we have introduced the function
\begin{equation}\label{deff}
    f_\epsilon(r,\theta)=\sum_{n=0}^{\infty}(-\sm i)^nJ_{n+\epsilon}(r)\sm e^{in\theta}.
\end{equation}
Consider its Laplace transform w.r.t.\ \(r\), given by
\begin{equation}\label{lapf}
   \mathcal{L}\sm\big[f_\epsilon(r,\theta)\big]=\frac{\big(s+\sqrt{1+s^2}\,\big)^{1-\epsilon}}{\sqrt{1+s^2}\,\big(s+\sqrt{1+s^2}+ie^{i\theta}\big)},
\end{equation}
\(\smash{\Re[s]>0}\). It follows from exploiting the linearity of the Laplace transform, the identity \cite[Sec.\ 17.13.103.]{GH}
\begin{equation}\label{Lap}
    \mathcal{L}\sm\big[J_\nu(r)\big]=\frac{\big(s+\sqrt{1+s^2}\big)^{-\sm\nu}}{\sqrt{1+s^2}},~ \nu>-1,~\Re[s]>0,
\end{equation}
and finally the geometric series formula \(\sum_{n\ge0}z^n=(1-z)^{-1}\) for \(|z|<1\). For half-integer flux, \(\smash{\lfloor\alpha\rfloor=n}\) and \(\smash{\{\alpha\}=1/2}\), the Laplace transform of \eqref{template2} reads
\begin{widetext}
\begin{align}\label{mouthful}
   \mathcal{L}\sm\big[F_{n+1/2}(r,\theta)\big]&=\sqrt{-\sm i}\,e^{-\sm in\theta}\sm\mathcal{L}\sm\big[f_{1/2}(r,\theta)\big]+\sqrt{-\sm i}\,e^{-\sm i(n+1)\theta}\sm\mathcal{L}\sm\big[f_{1/2}(r,-\sm\theta)\big]\nonumber\\[10pt]
   &\overset{\eqref{lapf}}{=}\sqrt{-\sm i}\,e^{-\sm in\theta}\frac{\sqrt{s+\sqrt{1+s^2}}}{\sqrt{1+s^2}}\left[\frac{\big(1+e^{-\sm i\theta}\big)\big(s+\sqrt{1+s^2}+i\big)}{\big(s+\sqrt{1+s^2}+ie^{i\theta}\big)\big(s+\sqrt{1+s^2}+ie^{-\sm i\theta}\big)}\right]\!.
\end{align}
Following careful simplifications, this reduces to
\begin{align}\label{prod}
    \mathcal{L}\sm\big[F_{n+1/2}(r,\theta)\big]
   =\frac{1}{2}\sm\sqrt{-\sm i}\,e^{-\sm in\theta}\big(1+e^{-\sm i\theta}\big)\frac{s+\sqrt{1+s^2}+i}{\sqrt{1+s^2}\,\sqrt{s+\sqrt{1+s^2}}}\cdot\frac{1}{s+i\cos\theta}.
\end{align}
\end{widetext}
Now, in order to invert the Laplace transform, note that 
\begin{equation}
    \mathcal{L}^{-1}\!\left[\frac{1}{s+i\cos\theta}\right]=e^{-\sm ir\cos\theta}\quad \text{\cite[Sec.\ 17.13.7.]{GH}},
\end{equation}
and
\begin{align}
    &\mathcal{L}^{-1}\!\left[\frac{s+\sqrt{1+s^2}+i}{\sqrt{1+s^2}\,\sqrt{s+\sqrt{1+s^2}}}\right]=\mathcal{L}^{-1}\!\left[\frac{\sqrt{s+\sqrt{1+s^2}}}{\sqrt{1+s^2}}\right]\nonumber\\
    &\quad + i\,\mathcal{L}^{-1}\!\left[\frac{1}{\sqrt{1+s^2}\,\sqrt{s+\sqrt{1+s^2}}}\right]=\sqrt{\frac{2}{\pi r}}\,e^{ir}\!,
\end{align}
given
\begin{align}
    \mathcal{L}^{-1}\!\left[\frac{\big(s+\sqrt{1+s^2}\sm\big)^{\pm1/2}}{\sqrt{1+s^2}}\right]\overset{\eqref{Lap}}{=}J_{\mp1/2}(r)=\sqrt{\frac{2}{\pi r}}\,\mqty{\displaystyle\cos\\[-3pt]\displaystyle\sin}\sm(r).
\end{align}
Applying the Laplace-convolution (or Faltung) theorem,% \cite[Sec.\ 17.12.5.]{GH},
\[\mathcal{L}^{-1}[f(s)\sm g(s)]=\mathcal{L}^{-1}[f(s)]*\mathcal{L}^{-1}[g(s)]\sm,\]
to \eqref{prod}, we are thus led to
\begin{equation}\label{nice}
    F_{n+1/2}(r,\theta)=\frac{e^{-\sm in\theta}}{\sqrt{2\sm\pi i}}\sm\big(1+e^{-\sm i\theta}\big)\int_0^r\frac{dr^\prime}{\sqrt{r^\prime}}\, e^{ir^\prime}\!e^{-\sm i(r-r^\prime)\cos\theta}\!.
\end{equation}
Observe that \eqref{nice} vanishes at \(\smash{\theta=\pi}\) due to the factor in parenthesis.  But for \(\smash{\theta\ne\pi}\), it follows that \(\smash{1+\cos\theta>0}\), hence substituting \(\smash{z=\sqrt{r^\prime(1+\cos\theta)}}\) in the above integral yields
\begin{align}
    F_{n+1/2}(r,\theta)&=\sqrt{\frac{2}{\pi i}}\,\frac{1+e^{-\sm i\theta}}{\sqrt{1+\cos\theta}}\, e^{-\sm in\theta-\sm ir\cos\theta}\nonumber\\
    &\qquad\times\!\int_0^{\sqrt{r\sm(1+\cos\theta)}}\!\!\!\!\!\!dz~ e^{iz^2}\!.
\end{align}
To complete the calculation, note that \(1+e^{-\sm i\theta}=2\sm e^{-\sm i\theta/2}\cos(\theta/2),\) and \(\smash{\sqrt{1+\cos\theta}=\sqrt{2}\,|\cos(\theta/2)|}\), which together imply the final result:
\begin{align}\label{finalres}
    F_{n+1/2}(r,\theta)&=\frac{2}{\sqrt{\pi i}}\,\text{sgn}\!\left(\cos\frac{\theta}{2}\right)e^{-\sm i(n+1/2)\theta-\sm ir\cos\theta}\nonumber\\
    &\kern2.5cm\times\!\int_0^{\sqrt{r\sm(1+\cos\theta)}}\!\!\!\!\!\!dz~ e^{iz^2}\!,
\end{align}
where \(\text{sgn}(\sm\cdot\sm)\) is the signum function. Clearly, \eqref{finalres} is a near duplicate of \eqref{AB23}, modulo \(\theta\)-dependent factors that had hitherto escaped notice. These insure that \eqref{finalres} is single-valued, unlike the function defined by \eqref{AB23}.

In any case, we emphasize that none of AB's {\emph {physical}} conclusions rely on \eqref{AB23}, and are therefore unaffected. This brief comment is meant in no way to denigrate AB's discovery, whose significance can hardly be overemphasised.

The author is  grateful to J.\ M.\ Wilkes for valuable editorial inputs.

\bibliography{AB}

%merlin.mbs apsrev4-1.bst 2010-07-25 4.21a (PWD, AO, DPC) hacked
%Control: key (0)
%Control: author (8) initials jnrlst
%Control: editor formatted (1) identically to author
%Control: production of article title (-1) disabled
%Control: page (0) single
%Control: year (1) truncated
%Control: production of eprint (0) enabled
\begin{thebibliography}{3}%
\makeatletter
\providecommand \@ifxundefined [1]{%
 \@ifx{#1\undefined}
}%
\providecommand \@ifnum [1]{%
 \ifnum #1\expandafter \@firstoftwo
 \else \expandafter \@secondoftwo
 \fi
}%
\providecommand \@ifx [1]{%
 \ifx #1\expandafter \@firstoftwo
 \else \expandafter \@secondoftwo
 \fi
}%
\providecommand \natexlab [1]{#1}%
\providecommand \enquote  [1]{``#1''}%
\providecommand \bibnamefont  [1]{#1}%
\providecommand \bibfnamefont [1]{#1}%
\providecommand \citenamefont [1]{#1}%
\providecommand \href@noop [0]{\@secondoftwo}%
\providecommand \href [0]{\begingroup \@sanitize@url \@href}%
\providecommand \@href[1]{\@@startlink{#1}\@@href}%
\providecommand \@@href[1]{\endgroup#1\@@endlink}%
\providecommand \@sanitize@url [0]{\catcode `\\12\catcode `\$12\catcode
  `\&12\catcode `\#12\catcode `\^12\catcode `\_12\catcode `\%12\relax}%
\providecommand \@@startlink[1]{}%
\providecommand \@@endlink[0]{}%
\providecommand \url  [0]{\begingroup\@sanitize@url \@url }%
\providecommand \@url [1]{\endgroup\@href {#1}{\urlprefix }}%
\providecommand \urlprefix  [0]{URL }%
\providecommand \Eprint [0]{\href }%
\providecommand \doibase [0]{http://dx.doi.org/}%
\providecommand \selectlanguage [0]{\@gobble}%
\providecommand \bibinfo  [0]{\@secondoftwo}%
\providecommand \bibfield  [0]{\@secondoftwo}%
\providecommand \translation [1]{[#1]}%
\providecommand \BibitemOpen [0]{}%
\providecommand \bibitemStop [0]{}%
\providecommand \bibitemNoStop [0]{.\EOS\space}%
\providecommand \EOS [0]{\spacefactor3000\relax}%
\providecommand \BibitemShut  [1]{\csname bibitem#1\endcsname}%
\let\auto@bib@innerbib\@empty
%</preamble>
\bibitem [{\citenamefont {Aharonov}\ and\ \citenamefont {Bohm}(1959)}]{AB59}%
  \BibitemOpen
  \bibfield  {author} {\bibinfo {author} {\bibfnamefont {Y.}~\bibnamefont
  {Aharonov}}\ and\ \bibinfo {author} {\bibfnamefont {D.}~\bibnamefont
  {Bohm}},\ }\href {\doibase 10.1103/PhysRev.115.485} {\bibfield  {journal}
  {\bibinfo  {journal} {Phys. Rev.}\ }\textbf {\bibinfo {volume} {115}},\
  \bibinfo {pages} {485} (\bibinfo {year} {1959})}\BibitemShut {NoStop}%
\bibitem [{\citenamefont {Kretzschmar}(1965)}]{propagator}%
  \BibitemOpen
  \bibfield  {author} {\bibinfo {author} {\bibfnamefont {M.}~\bibnamefont
  {Kretzschmar}},\ }\href {\doibase https://doi.org/10.1007/BF01381305}
  {\bibfield  {journal} {\bibinfo  {journal} {Z. Physik}\ }\textbf {\bibinfo
  {volume} {185}},\ \bibinfo {pages} {84} (\bibinfo {year} {1965})}\BibitemShut
  {NoStop}%
\bibitem [{\citenamefont {Gradshteyn}\ and\ \citenamefont {Ryzhik}(2007)}]{GH}%
  \BibitemOpen
  \bibfield  {author} {\bibinfo {author} {\bibfnamefont {I.~S.}\ \bibnamefont
  {Gradshteyn}}\ and\ \bibinfo {author} {\bibfnamefont {I.~M.}\ \bibnamefont
  {Ryzhik}},\ }\href {http://fisica.ciens.ucv.ve/~svincenz/TISPISGIMR.pdf}
  {\emph {\bibinfo {title} {Table of integrals, series, and products}}},\
  \bibinfo {edition} {4th}\ ed.\ (\bibinfo  {publisher} {Elsevier},\ \bibinfo
  {address} {New York},\ \bibinfo {year} {2007})\BibitemShut {NoStop}%
\end{thebibliography}%
\end{document}